\documentclass[12pt]{article}
\usepackage[margin=1.9cm]{geometry}
\usepackage{epsfig}
\usepackage{cite}

\newcommand{\mysection}{\setcounter{equation}{0}\section}

\def\beq{\begin{equation}}
\def\eeq{\end{equation}}
\def\beqa{\begin{eqnarray}}
\def\eeqa{\end{eqnarray}}
 
\begin{document}

\begin{center}
{\Large \bf Soft-gluon corrections for the associated production of a single top quark and a Higgs boson}
\end{center}
\vspace{2mm}
\begin{center}
{\large Matthew Forslund$^a$ and Nikolaos Kidonakis$^b$}\\
\vspace{2mm}
${}^a${\it Department of Physics and Astronomy, Stony Brook University, \\ 
Stony Brook, NY 11794, USA}

\vspace{1mm}

${}^b${\it Department of Physics, Kennesaw State University, \\
Kennesaw, GA 30144, USA}
\end{center}

\begin{abstract}
  We discuss soft-gluon resummation for the associated production of a single top quark and a Higgs boson ($tqH$ production) in single-particle-inclusive (1PI) kinematics. We present analytical results for the higher-order corrections and numerical results for the cross sections at LHC energies. We calculate approximate NNLO total rates, including scale dependence and uncertainties from parton distributions, as well as top-quark transverse-momentum and rapidity distributions. In all cases we find that the soft-gluon corrections are dominant and provide important contributions to the total and differential cross sections.
\end{abstract}

\mysection{Introduction}

The top quark and the Higgs boson occupy a central role in particle-physics studies at current collider energies. The production of a single top quark in association with a Higgs boson is a very interesting process as it involves both of these very massive particles, and its cross section is sensitive to top-quark and Higgs couplings. Thus, modifications to the couplings due to new physics beyond the Standard Model would directly affect the cross section for $tqH$ production. Searches for this process have been made at the LHC at 8 TeV \cite{CMS8tev} and 13 TeV \cite{CMS13tev} energies. Next-to-leading-order (NLO) calculations for $tqH$ production have been done in Refs. \cite{CER,DMMZ}. Given the large size of the NLO corrections, it is important to calculate radiative corrections beyond that order.

Soft-gluon resummation is a powerful formalism for making theoretical predictions for perturbative corrections at higher orders. The resummation is a consequence of the factorization properties of the cross section \cite{NKGS,CLS,KOS,LOS,HRSV,FK2020}. The soft-gluon corrections in the perturbative series are in the form of plus distributions that involve logarithms of a threshold variable that measures the energy in the soft emission.

For many processes, and in particular for top-quark production (see Ref. \cite{NKtoprev} for a review), these soft-gluon corrections are large and they numerically dominate the corrections at higher orders. Thus, they can be considered as excellent approximations to complete results at NLO, next-to-NLO (NNLO), etc. Specific processes for which soft-gluon corrections have been known to provide excellent approximations are top-antitop pair production \cite{NKn3lo}, single-top production in the $s$, $t$, and $tW$ channels \cite{NKst,NKNY}, top production in association with a charged Higgs boson \cite{NKtH}, and top production via anomalous couplings in association with a $Z$ boson \cite{NKtZ}, a photon \cite{MFNK}, or a $Z'$ boson \cite{MGNK}. As we will show in this paper (see also \cite{FK2020}), $tqH$ production is another such process.

The choice of threshold variable in the resummation depends on the kinematics. While most calculations in the past have been done for $2 \to 2$ processes in single-particle-inclusive (1PI) kinematics or pair-invariant-mass kinematics (see the reviews in Refs. \cite{NKtoprev,NK2020rev}), and some additional work has been done on $2 \to 3$ processes such as $t{\bar t}X$ \cite{LLL,KMST,BFPSY,BFOP,BFPY,BFOPS,KMST2,KMSST,BFFPPT} using three-particle-invariant-mass kinematics, the resummation formalism has been recently extended in 1PI kinematics \cite{FK2020} to processes with an arbitrary number of final-state particles. 

In this paper, we employ our resummation formalism in \cite{FK2020} to calculate cross sections for $tqH$ production. In addition, our formalism also allows the calculation of 1PI differential distributions in transverse momentum and rapidity. We begin in Section 2 with a description of soft-gluon resummation for $tqH$ production. We define a threshold variable $s_4$ that measures the additional energy in soft radiation and that vanishes at partonic threshold, and we derive the exponentiation of the logarithms of that threshold variable. We also present results for the soft anomalous dimensions through two loops, and discuss fixed-order expansions. In Section 3, numerical results are presented for $tqH+{\bar t}qH$ production at LHC energies, including total cross sections as well as top-quark transverse-momentum and rapidity distributions. We conclude in Section 4.

\mysection{Resummation for $tqH$ production}

\begin{figure}[htbp]
\begin{center}
\includegraphics[width=12cm]{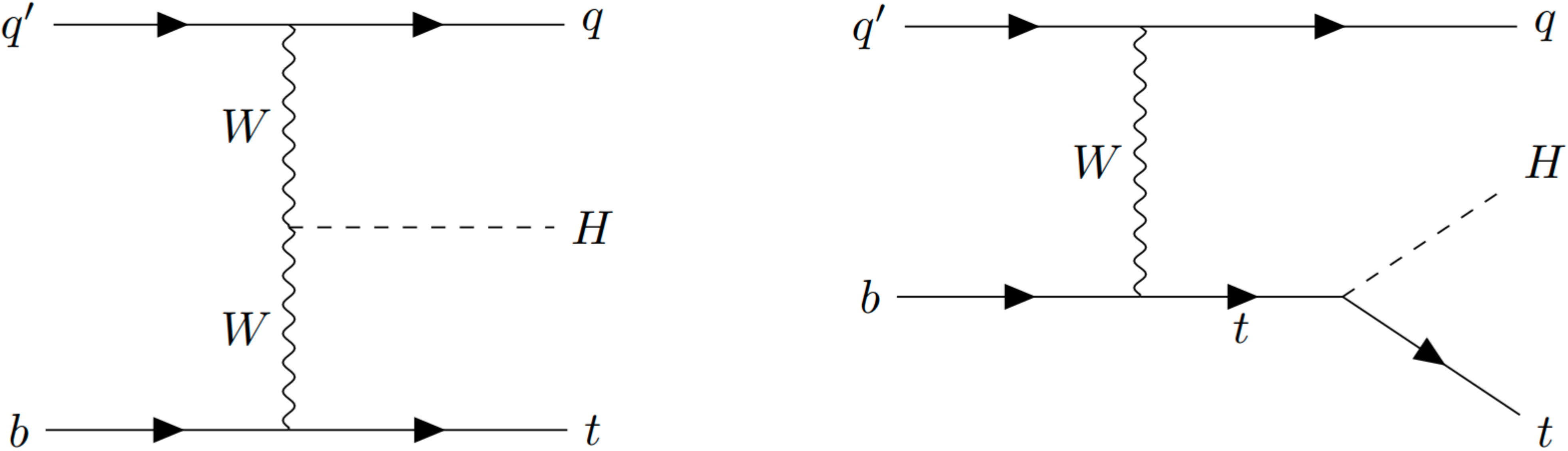}
\caption{Leading-order diagrams for $tqH$ production.}
\label{diagrams}
\end{center}
\end{figure}

In this section we review the formalism for soft-gluon resummation in 1PI kinematics for $tqH$ production. We consider the partonic process $a(p_a)+b(p_b) \to t(p_1)+q(p_2)+H(p_3)$ (i.e., $bq' \to tqH$). The leading-order diagrams are shown in Fig. \ref{diagrams}.
We define the parton-level kinematical variables $s=(p_a+p_b)^2$, $t=(p_a-p_1)^2$, and $u=(p_b-p_1)^2$. If we have emission of an additional gluon with momentum $p_g$ in the final state, then momentum conservation requires that $p_a +p_b=p_1 +p_2 +p_3+p_g$.

We then define a threshold variable $s_4=(p_2+p_3+p_g)^2-(p_2+p_3)^2$. This variable describes the extra energy from gluon emission and clearly vanishes as $p_g \to 0$. It is straightforward to show that the above expression is equivalent to $s_4=s+t+u-p_1^2-p_{23}^2$ where we have defined $p_{23}=p_2+p_3$.

If the initial-state partons $a$ and $b$ are from protons $A$ and $B$, we define the hadron-level kinematical variables $S=(p_A+p_B)^2$, $T=(p_A-p_1)^2$, and $U=(p_B-p_1)^2$, and note that $p_a=x_a \, p_A$ and $p_b=x_b \, p_B$ with $x_a$, $x_b$ momentum fractions for the partons. We also define the variable $S_4=S+T+U-p_1^2-p_{23}^2$ and, after some algebra, we derive the relation 
\beq
\frac{S_4}{S}=\frac{s_4}{s}-(1-x_a)\frac{\left(u-p_{23}^2\right)}{s}-(1-x_b)\frac{\left(t-p_{23}^2\right)}{s}
+(1-x_a)(1-x_b) \frac{\left(p_1^2-p_{23}^2\right)}{s} \, .
\label{s4}
\eeq
The last term in the above equation involves $(1-x_a)(1-x_b)$ and can, thus, be ignored in the threshold limit, $x_a \to 1$ and $x_b \to 1$.

\subsection{Resummed cross section}

We write the differential cross section in 1PI kinematics for $tqH$ production in proton-proton collisions as a convolution, 
\beq
E_1\frac{d\sigma_{pp \to tqH}}{d^3p_1}=\sum_{a,b} \; 
\int dx_a \, dx_b \,  \phi_{a/A}(x_a, \mu_F) \, \phi_{b/B}(x_b, \mu_F) \, 
E_1 \frac{d{\hat \sigma}_{ab \to tqH}(s_4, \mu_F)}{d^3p_1} \, ,
\label{facphi}
\eeq
where $E_1$ is the energy of the observed top quark, $\phi_{a/A}$  and $\phi_{b/B}$ are the parton distribution functions (pdf) for partons $a$ and $b$ in protons $A$ and $B$, respectively, ${\hat \sigma}_{ab \to tqH}$ is the hard-scattering partonic cross section, and $\mu_F$ is the factorization scale. The cross section at each perturbative order also depends on the renormalization scale, $\mu_R$.

The cross section factorizes in integral transform space \cite{NKGS,CLS,LOS,KOS,HRSV,FK2020}. We define Laplace transforms, denoted below by a tilde, of the partonic cross section as ${\tilde{\hat\sigma}}_{ab \to tqH}(N)=\int_0^s (ds_4/s) \,  e^{-N s_4/s} \, {\hat\sigma}_{ab \to tqH}(s_4)$, where $N$ is the transform variable. The logarithms of $s_4$ appearing in the perturbative series transform into logarithms of $N$ and, as we will see, the latter exponentiate. We also define transforms of the pdf via ${\tilde \phi}(N)=\int_0^1 e^{-N(1-x)} \phi(x) \, dx$. 

We next write the parton-parton cross section, $E_1 \, d\sigma_{ab \to tqH}/d^3p_1$, in the same form as Eq. (\ref{facphi}) but with the initial-state hadrons replaced by partons \cite{NKGS,CLS,LOS,KOS,HRSV,FK2020}  
\beq
E_1\frac{d\sigma_{ab \to tqH}(S_4)}{d^3p_1}=
\int dx_a \, dx_b \,  \phi_{a/a}(x_a) \, \phi_{b/b}(x_b) \, 
E_1 \frac{d{\hat \sigma}_{ab \to tqH}(s_4)}{d^3p_1} \, ,
\label{facpaphi}
\eeq
where for simplicity we suppress the dependence on the scales, and we define its transform as 
\beq
E_1\frac{d{\tilde \sigma}_{ab \to tqH}(N)}{d^3p_1}=\int_0^S 
\frac{dS_4}{S} \,  e^{-N S_4/S} \, E_1\frac{d\sigma_{a b \to tqH}(S_4)}{d^3p_1} \, . 
\label{csmom}
\eeq 

By taking a transform of Eq. (\ref{facpaphi}), as defined in Eq. (\ref{csmom}), and taking into consideration Eq. (\ref{s4}), we have  
\beqa
E_1 \frac{d{\tilde \sigma}_{ab \to tqH}(N)}{d^3p_1} &=& \int_0^1 dx_a e^{-N_a (1-x_a)} \phi_{a/a}(x_a) \int_0^1 dx_b e^{-N_b (1-x_b)} \phi_{b/b}(x_b)
\nonumber \\ && \times
\int_0^s \frac{ds_4}{s} e^{-N s_4/s} E_1 \frac{d{\hat \sigma}_{ab \to tqH}(s_4)}{d^3p_1}
\nonumber \\ 
&=& {\tilde \phi}_{a/a}(N_a) \, {\tilde \phi}_{b/b}(N_b) \, E_1 \frac{d{\tilde{\hat \sigma}}_{ab \to tqH}(N)}{d^3p_1} \, ,
\label{fac}
\eeqa
where $N_a=N(p^2_{23}-u)/s$ and $N_b=N(p^2_{23}-t)/s$.

We then provide a new refactorized form of the cross section in terms of a different set of functions \cite{NKGS,CLS,LOS,KOS,HRSV,FK2020}. We first rewrite Eq. (\ref{s4}) in the form 
\beqa
\frac{S_4}{S}& = & -(1-x_a)\frac{\left(u-p^2_{23}\right)}{s}-(1-x_b)\frac{\left(t-p^2_{23}\right)}{s}+\frac{s_4}{s}
\nonumber \\ 
&=& -w_a \frac{(u-p^2_{23})}{s}- w_b \frac{(t-p^2_{23})}{s}+w_S +w_q \, ,
\label{ws}
\eeqa
where the $w$'s denote dimensionless weights, and $w_a \neq 1-x_a$ and $w_b \neq 1-x_b$ because they refer to different functions.
Then, we find a refactorized form for the cross section \cite{NKGS,LOS,FK2020} as 
\beqa
E_1\frac{d{\sigma}_{ab \to tqH}}{d^3p_1}&=&\int dw_a \, dw_b \, dw_q \, dw_S \, \psi_{a/a}(w_a) \, \psi_{b/b}(w_b) \, J_q(w_q) 
\nonumber \\ && \hspace{-30mm} \times 
{\rm tr} \left\{H_{ab \to tqH}\left(\alpha_s(\mu_R)\right) \, 
S_{ab \to tqH}\left(\frac{w_S \sqrt{s}}{\mu_F} \right)\right\} \; 
\delta\left(\frac{S_4}{S}+w_a\frac{(u-p^2_{23})}{s}
+w_b \frac{(t-p^2_{23})}{s}-w_S -w_q\right) \, .
\nonumber \\ 
\label{refact}
\eeqa
The hard function, $H_{ab \to tqH}$, is infrared safe and it receives contributions from the amplitude and its complex conjugate for the process, while    
the soft function, $S_{ab \to tqH}$, describes the emission of noncollinear soft gluons. The hard and the soft functions are both matrices in the color space of the partonic scattering ($2\times 2$ matrices for $bq' \to tqH$) and, thus, we take the trace of their product in the above equation. 
The functions $\psi$ are distributions for incoming partons at fixed value of momentum and involve collinear emission, and they differ from the pdf $\phi$ \cite{NKGS,CLS,KOS,LOS,HRSV,FK2020,GS}. The function $J_q$ describes collinear emission from the final-state light quark.

After taking a transform of Eq. (\ref{refact}) via Eq. (\ref{csmom}), and using Eq. (\ref{ws}), we then find
\beqa
E_1\frac{d{\tilde \sigma}_{ab \to tqH}(N)}{d^3p_1}&=& 
\int_0^1 dw_a e^{-N_a w_a} \psi_{a/a}(w_a) \int_0^1 dw_b e^{-N_b w_b} \psi_{b/b}(w_b)
\nonumber \\ && \hspace{-25mm} \times
\int_0^1 dw_q e^{-N w_q} J_q(w_q) \; \,
{\rm tr}\left\{H_{ab \to tqH}\left(\alpha_s(\mu_R)\right) \int_0^1 dw_s e^{-N w_s}  
S_{ab \to tqH}\left(\frac{w_s\sqrt{s}}{\mu_F} \right)\right\} 
\nonumber \\ && \hspace{-25mm} 
={\tilde \psi}_{a/a}(N_a) \, {\tilde \psi}_{b/b}(N_b) {\tilde J_q}\left(N \right) \; \,
{\rm tr} \left\{H_{ab \to tqH}\left(\alpha_s(\mu_R)\right) \, 
{\tilde S}_{ab \to tqH}\left(\frac{\sqrt{s}}{N \mu_F} \right)\right\} \, .
\label{refac}
\eeqa
We see that the hard function $H_{ab \to tqH}$ is independent of $N$, and all the $N$-dependence is absorbed in the functions ${\tilde S}_{ab \to tqH}$, ${\tilde \psi}$, and ${\tilde J_q}$, in contrast to the original form where both the partonic cross section ${\tilde {\hat \sigma}}$ and the parton densities ${\tilde \phi}$ have $N$-dependence \cite{NKGS,CLS,KOS,LOS,HRSV,FK2020,GS}.

By comparing Eq. (\ref{fac}) with Eq. (\ref{refac}), we find an expression for the hard-scattering partonic cross section in transform space,
\beq
E_1\frac{d{\tilde{\hat \sigma}}_{ab \to tqH}(N)}{d^3p_1}=
\frac{{\tilde \psi}_a(N_a) \, {\tilde \psi}_b(N_b) \, {\tilde J_q} (N)}{{\tilde \phi}_{a/a}(N_a) \, 
{\tilde \phi}_{b/b}(N_b)} \; \,  {\rm tr} \left\{H_{ab \to tqH}\left(\alpha_s(\mu_R)\right) \, 
{\tilde S}_{ab \to tqH}\left(\frac{\sqrt{s}}{N \mu_F} \right)\right\} \, .
\label{sigN}
\eeq

The dependence of the soft matrix, ${\tilde S}_{ab \to tqH}$, on the transform variable, $N$, is resummed via renormalization-group evolution \cite{NKGS}. We have 
\beq
{\tilde S}^b_{ab \to tqH}=(Z_{\!S\, ab \to tqH})^{\dagger} \; {\tilde S}_{ab \to tqH} \; 
Z_{\!S\, ab \to tqH} \, ,
\eeq
where ${\tilde S}^b_{ab \to tqH}$ is the unrenormalized (bare) quantity 
and $Z_{\!S \, ab \to tqH}$ is a matrix of renormalization constants.
Thus, we find that ${\tilde S}_{ab \to tqH}$ obeys the renormalization-group equation
\beq
\left(\mu_R \frac{\partial}{\partial \mu_R}
+\beta(g_s)\frac{\partial}{\partial g_s}\right) {\tilde S}_{ab \to tqH}
=-(\Gamma_{\! S \, ab \to tqH})^{\dagger} \; {\tilde S}_{ab \to tqH}-{\tilde S}_{ab \to tqH} \; \Gamma_{\! S \, ab \to tqH}
\eeq
where $g_s^2=4\pi\alpha_s$ and $\beta$ is the QCD beta function. The soft anomalous dimension matrix, $\Gamma_{\! S \, ab \to tqH}$, controls the evolution of the soft function. Soft anomalous dimensions are calculated from the coefficients of the ultraviolet poles of the relevant eikonal diagrams \cite{NKGS,KOS,FK2020,NKst,ADS,NK2loop,NK3loop}.

The $N$-space resummed cross section is derived from the renormalization-group evolution of the $N$-dependent functions in Eq. (\ref{sigN}), i.e. ${\tilde S}_{ab \to tqH}$, ${\tilde \psi}$, ${\tilde \phi}$, and ${\tilde J}_q$, 
and it is given by \cite{NKGS,LOS,NKtoprev,HRSV,FK2020}
\beqa
E_1\frac{d{\tilde{\hat \sigma}}_{ab \to tqH}^{\rm resum}(N)}{d^3p_1} &=&
\exp\left[\sum_{i=a,b} E_{i}(N_i)\right] \, 
\exp\left[\sum_{i=a,b} 2 \int_{\mu_F}^{\sqrt{s}} \frac{d\mu}{\mu} \gamma_{i/i}(N_i)\right] \, 
\exp\left[E'_q(N)\right]
\nonumber\\ && \hspace{-5mm} \times \,
{\rm tr} \left\{H_{ab \to tqH}\left(\alpha_s(\sqrt{s})\right) {\bar P} \exp \left[\int_{\sqrt{s}}^{{\sqrt{s}}/N}
\frac{d\mu}{\mu} \; \Gamma_{\! S \, ab \to tqH}^{\dagger} \left(\alpha_s(\mu)\right)\right] \; \right.
\nonumber\\ && \left. \hspace{5mm} \times \,
{\tilde S}_{ab \to tqH} \left(\alpha_s\left(\frac{\sqrt{s}}{N}\right)\right) \;
P \exp \left[\int_{\sqrt{s}}^{{\sqrt{s}}/N}
\frac{d\mu}{\mu}\; \Gamma_{\! S \, ab \to tqH}
\left(\alpha_s(\mu)\right)\right] \right\} \, ,
\nonumber \\
\label{resummed}
\eeqa
where $P$ denotes path-ordering in the same sense as the integration variable $\mu$, and ${\bar P}$ denotes path-ordering in the reverse sense. This moment-space resummed cross section resums logarithms of the moment variable $N$. For next-to-leading-logarithm (NLL) resummation we need one-loop calculations for the process-dependent soft anomalous dimensions, while for next-to-NLL (NNLL) resummation we need two-loop calculations. 

The first exponential in Eq. (\ref{resummed}) resums soft and collinear emission from the initial-state partons \cite{GS,CT}, 
\beq
E_i(N_i)=
\int^1_0 dz \frac{z^{N_i-1}-1}{1-z}\;
\left \{\int_1^{(1-z)^2} \frac{d\lambda}{\lambda}
A_i\left(\alpha_s(\lambda s)\right)
+D_i\left[\alpha_s((1-z)^2 s)\right]\right\} \, ,
\label{Eexp}
\eeq
where $A_i = \sum_{k=1}^{\infty} (\alpha_s/\pi)^k A_i^{(k)}$. We have  
$A_i^{(1)}=C_i$ with $C_i=C_F=(N_c^2-1)/(2N_c)$ for a quark 
or antiquark and $C_i=C_A=N_c$ for a gluon, with $N_c=3$ the number of colors. 
Furthermore, $A_i^{(2)}=C_i K/2$ where $K= C_A\; ( 67/18-\pi^2/6 ) - 5n_f/9$,     
with $n_f$ the number of light quark flavors. 
We also have $D_i=\sum_{k=1}^{\infty} (\alpha_s/\pi)^k D_i^{(k)}$, 
with $D_i^{(1)}=0$.

The second exponential in Eq. (\ref{resummed}) provides the scale evolution in terms of the parton anomalous dimensions $\gamma_{i/i}=-A_i \ln N_i+\gamma_i$.  
Here $\gamma_i=\sum_{k=1}^{\infty} (\alpha_s/\pi)^k \gamma_i^{(k)}$, with $\gamma_q^{(1)}=3C_F/4$ for quarks and $\gamma_g^{(1)}=\beta_0/4$ for gluons, where $\beta_0=(11 C_A-2n_f)/3$.

The third exponential in Eq. (\ref{resummed}) describes radiation from the final-state quark \cite{KOS,LOS,GS,CT}. We have
\beqa
E'_q(N)&=&
\int^1_0 dz \frac{z^{N-1}-1}{1-z}\;
\left \{\int^{1-z}_{(1-z)^2} \frac{d\lambda}{\lambda}
A_q \left(\alpha_s\left(\lambda s\right)\right)
+B_q\left[\alpha_s((1-z)s)\right]
+D_q\left[\alpha_s((1-z)^2 s)\right]\right\} \, ,
\nonumber \\
\label{E'exp}
\eeqa
where $B_q=\sum_{k=1}^{\infty} (\alpha_s/\pi)^k B_q^{(k)}$, 
with $B_q^{(1)}=-3C_F/4$.

The process-dependent hard and soft functions, which are matrices, have the perturbative expansions 
$H_{ab \to tqH}=\sum_{k=0}^{\infty}(\alpha_s/\pi)^k \,  H_{ab \to tqH}^{(k)}$ 
and ${\tilde S}_{ab \to tqH}=\sum_{k=0}^{\infty} (\alpha_s/\pi)^k \, {\tilde S}_{ab \to tqH}^{(k)}$. Finally, the soft anomalous dimension matrix has the expansion $\Gamma_{\! S \, ab \to tqH}=\sum_{k=1}^{\infty}(\alpha_s/\pi)^k \, \Gamma_{\! S \, ab \to tqH}^{(k)}$.

\subsection{Soft anomalous dimension matrices}

Next, we present the soft anomalous dimension matrices for the processes 
$b(p_a)+q'(p_b) \rightarrow t(p_1) +q(p_2)+H(p_3)$ at one and two loops \cite{FK2020}.  
In addition to $s$, $t$, $u$ we also define the kinematical variables $s'=(p_1+p_2)^2$, $t'=(p_b-p_2)^2$, and $u'=(p_a-p_2)^2$, and we   
choose the color basis $c_1=\delta_{a1} \delta_{b2}$ and 
$c_2=T^c_{1a} T^c_{2b}$.
The four elements of the soft anomalous dimension matrix at one loop, where the $ij$th element is denoted by
$\Gamma_{\! S \, bq' \to tqH}^{ij \, (1)}$, are given in this color basis by 
\beqa
{\Gamma}_{\! S \, bq' \to tqH}^{11\, (1)}&=&
C_F \left[\ln\left(\frac{t'(t-m_t^2)}{m_t \, s^{3/2}}\right)-\frac{1}{2}\right] \, ,
\nonumber \\
{\Gamma}_{\! S \, bq' \to tqH}^{12 \, (1)}&=&\frac{C_F}{2N_c} \ln\left(\frac{u'(u-m_t^2)}{s(s'-m_t^2)}\right) \, ,
\quad
{\Gamma}_{\! S \, bq' \to tqH}^{21 \,(1)}= \ln\left(\frac{u'(u-m_t^2)}{s(s'-m_t^2)}\right) \, ,
\nonumber \\
{\Gamma}_{\! S \, bq' \to tqH}^{22 \, (1)}&=& C_F \left[\ln\left(\frac{t'(t-m_t^2)}{m_t \, s^{3/2}}\right)-\frac{1}{2}\right]
-\frac{1}{N_c}\ln\left(\frac{u'(u-m_t^2)}{s(s'-m_t^2)}\right) 
+\frac{N_c}{2}\ln\left(\frac{u'(u-m_t^2)}{t'(t-m_t^2)}\right) \, ,
\nonumber \\
\label{Gamma1}
\eeqa
where $m_t$ is the top-quark mass.

At two loops, the four matrix elements of the soft anomalous dimension, with the $ij$th element denoted by
$\Gamma_{\! S\, bq' \to tqH}^{ij \, (2)}$, can be written directly in terms of the one-loop results.
We have 
\beqa
\Gamma_{\! S\, bq' \to tqH}^{11 \,(2)}&=& \frac{K}{2} \Gamma_{\! S \, bq' \to tqH}^{11 \,(1)}+\frac{1}{4} C_F C_A (1-\zeta_3) \, ,
\quad
\Gamma_{\! S \, bq' \to tqH}^{12 \,(2)}= \frac{K}{2} \Gamma_{\! S \, bq' \to tqH}^{12 \, (1)} \, ,
\nonumber \\
\Gamma_{\! S \, bq' \to tqH}^{21 \,(2)}&=& \frac{K}{2} \Gamma_{\! S \, bq' \to tqH}^{21 \, (1)} \, ,
\quad
\Gamma_{\! S \, bq' \to tqH}^{22 \,(2)}= \frac{K}{2} \Gamma_{\! S \, bq' \to tqH}^{22 \,(1)}+\frac{1}{4} C_F C_A (1-\zeta_3) \, .
\label{Gamma2}
\eeqa

\subsection{Fixed-order expansions}

When we expand the resummed cross section, Eq. (\ref{resummed}), to a fixed order and then invert back to momentum space, we get powers of logarithms of $s_4$ in the form of plus distributions.

The NLO soft-gluon corrections in the cross section are then given by 
\beqa
E_1\frac{d{\hat{\sigma}}_{bq' \to tqH}^{(1)}}{d^3p_1}&=& 
F^{LO}_{bq' \to tqH} \frac{\alpha_s(\mu_R)}{\pi}
\left\{c_3^{bq' \to tqH} \, {\cal D}_1(s_4) + c_2^{bq' \to tqH} \, {\cal D}_0(s_4) 
+c_1^{bq' \to tqH} \, \delta(s_4)\right\}
\nonumber \\ &&
{}+\frac{\alpha_s(\mu_R)}{\pi} \, 
A_{bq' \to tqH} \, {\cal D}_0(s_4) \, ,
\label{aNLO}
\eeqa
where the plus distributions of the logarithms of $s_4$ are denoted by
${\cal D}_k(s_4)=[(\ln^k(s_4/m_t^2))/s_4]_+$,  
and $F^{LO}_{bq' \to tqH}= {\rm tr}\{H_{bq' \to tqH}^{(0)} \, S_{bq' \to tqH}^{(0)}\}$ involves the leading-order (LO) cross section. 
As can be seen from the above expression, only a part of the corrections is proportional to $F^{LO}_{bq' \to tqH}$. 

The coefficient of the leading logarithm at NLO, ${\cal D}_1(s_4)$, is given by 
$c_3^{bq' \to tqH}=3 \, C_F$.  
The coefficients of the next-to-leading logarithm at NLO, ${\cal D}_0(s_4)$, are given by
\beq
c_2^{bq' \to tqH}=-2 \, C_F \, \ln\left(\frac{(p^2_{23}-u)(p^2_{23}-t)}{m_t^4}\right) 
-\frac{3}{4}C_F -3 \, C_F \ln\left(\frac{m_t^2}{s}\right)
-2 \, C_F \ln\left(\frac{\mu_F^2}{m_t^2}\right)
\label{c2}
\eeq
and 
\beq
A_{bq' \to tqH}={\rm tr} \left\{H_{bq' \to tqH}^{(0)} \, \Gamma_{\! S \, bq' \to tqH}^{(1) \, \dagger} \, S_{bq' \to tqH}^{(0)}
+H_{bq' \to tqH}^{(0)} \, S_{bq' \to tqH}^{(0)} \, \Gamma_{\! S \, bq' \to tqH}^{(1)}\right\} \, .
\label{Atqh}
\eeq
We also have 
\beq
c_1^{bq' \to tqH}=C_F\left[\ln\left(\frac{(p^2_{23}-u)(p^2_{23}-t)}{m_t^4}\right)
-\frac{3}{2}\right]\ln\left(\frac{\mu_F^2}{m_t^2}\right)  \, .
\label{c1mu}
\eeq

The analytical form of the NNLO soft-gluon corrections is much longer and can be found from the expressions in \cite{FK2020}.

\mysection{Cross sections for $tqH+{\bar t}qH$ production at LHC energies}

In this section we present results for the production cross section for $tqH+{\bar t}qH$ as well as top-quark differential distributions in transverse momentum and rapidity. We set the top-quark mass $m_t=172.5$ GeV and the Higgs-boson mass $m_H=125$ GeV, and we use the latest MSHT20 \cite{MSHT20} and CT18 \cite{CT18} pdf sets via LHAPDF6 \cite{lhapdf}. The calculations of the cross sections at each perturbative order use the pdf provided at that order.  We set the factorization and renormalization scales equal to each other and denote this common scale by $\mu$. We compute higher-order soft-gluon corrections from resummation at next-to-leading-logarithm (NLL) accuracy. The complete NLO results are found by using MadGraph5\_aMC@NLO \cite{MG5}. In our discussion, we denote the sum of the complete NLO cross section and the NNLO soft-gluon corrections as approximate NNLO (aNNLO).

\subsection{Total cross sections}

\begin{figure}[htbp]
\begin{center}
\includegraphics[width=91mm]{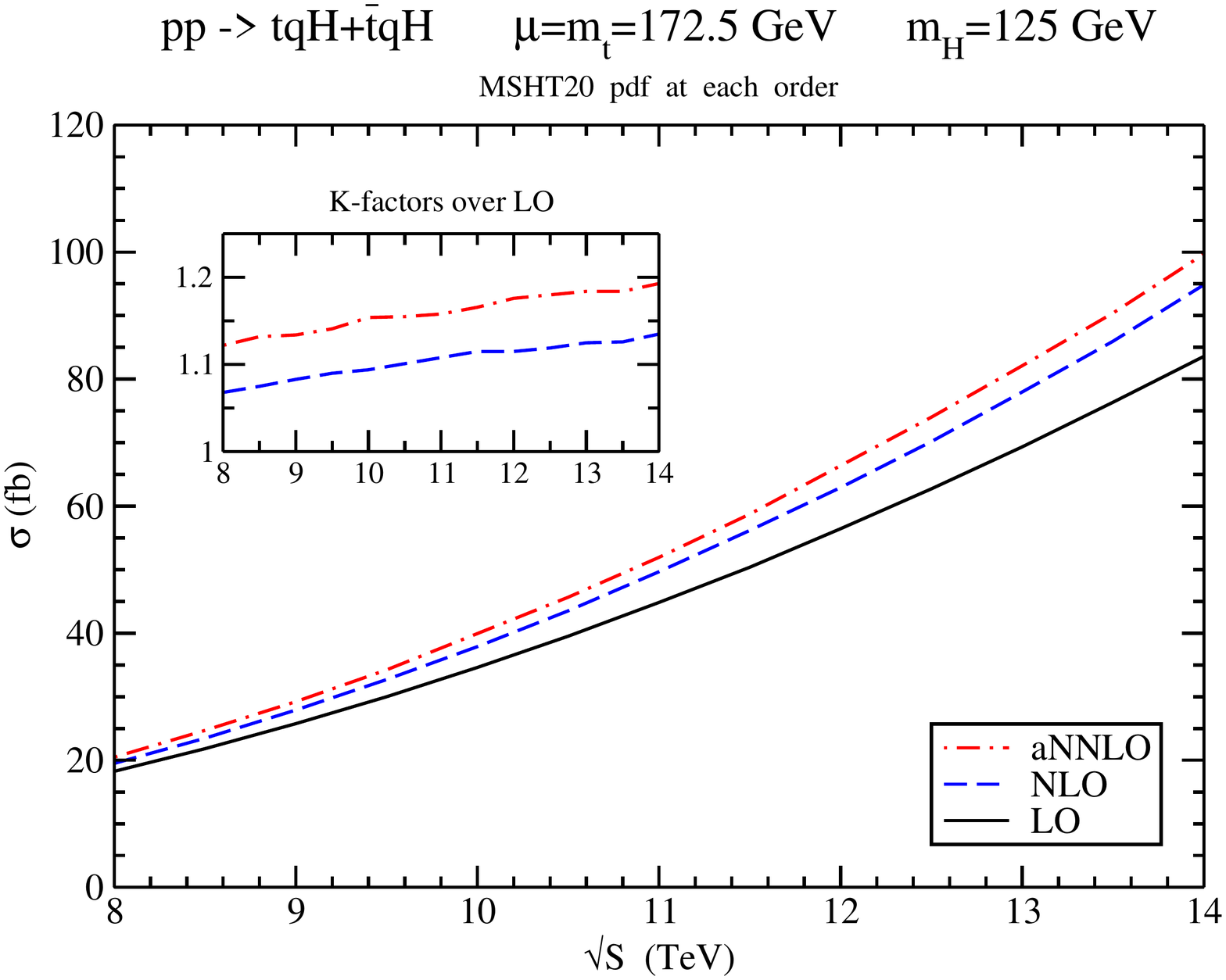}
\hspace{-7mm}
\includegraphics[width=91mm]{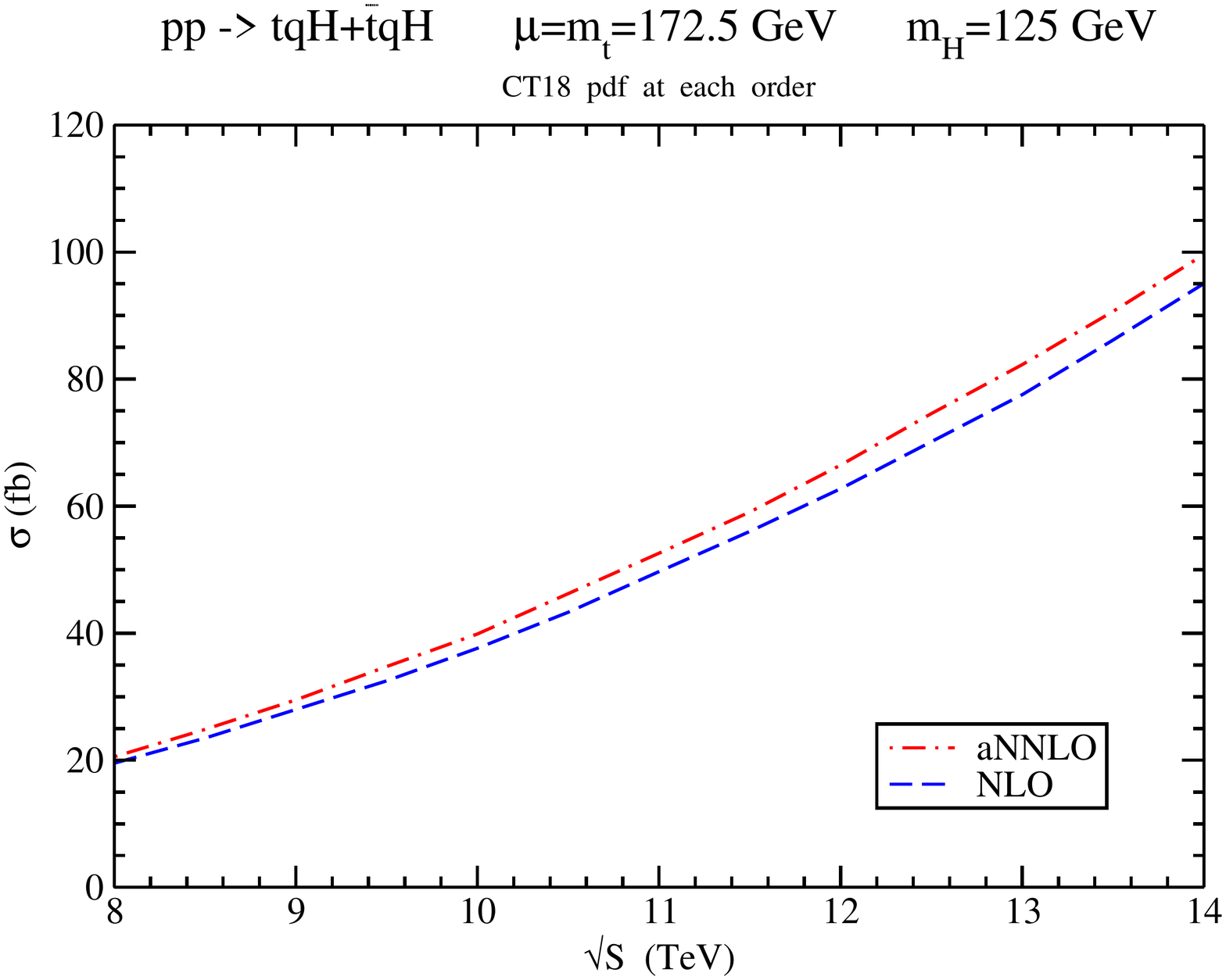}
\caption{The total cross sections at LO, NLO, and aNNLO for $tqH+{\bar t}qH$ production at LHC energies using (left) MSHT20 pdf and (right) CT18 pdf.}
\label{tqHrootS}
\end{center}
\end{figure}

In Fig. \ref{tqHrootS} we display the total cross sections with $\mu=m_t$ for LHC energies from 8 to 14 TeV. The plot on the left shows the LO cross section using MSHT20 LO pdf, the NLO cross section using MSHT20 NLO pdf, and the aNNLO cross section using MSHT20 NNLO pdf. The inset plot on the left displays the $K$-factors, i.e. the ratios of the NLO and aNNLO cross sections to the LO cross section. We see that the NLO corrections are important and increase with energy, and the further aNNLO corrections are quite significant: the NLO/LO $K$-factor indicates corrections of nearly 13.5\% at 14 TeV energy while the aNNLO/LO $K$-factor indicates corrections of nearly 19.3\% at 14 TeV energy. The plot on the right of Fig. \ref{tqHrootS} shows the NLO cross section using CT18 NLO pdf and the aNNLO cross section using CT18 NNLO pdf (we do not show LO results with CT18 pdf as there are no LO pdf provided by that set). The results with CT18 pdf are almost identical to those with MSHT20 pdf. 

\begin{table}[htb]
\begin{center}
\begin{tabular}{|c|c|c|c|c|} \hline
\multicolumn{5}{|c|}{$tqH+{\bar t}qH$ cross sections} \\ \hline
$\sigma$ in fb & LO & aNLO & NLO & aNNLO \\ \hline
LHC 8 TeV  & 18.2 & 19.5 & 19.5 & 20.5  \\ \hline
LHC 13 TeV & 69.2 & 80.4 & 78.0 & 81.9 \\ \hline
LHC 14 TeV & 83.4 & 98.2 & 94.8 & 99.5 \\ \hline
\end{tabular}
\caption[]{The $tqH+{\bar t}qH$ cross sections (in fb, with $\mu=m_t$) in $pp$ collisions at the LHC with $\sqrt{S}=8$, 13, and 14 TeV, with $m_t=172.5$ GeV, $m_H=125$ GeV, and MSHT20 pdf at each order.}
\label{table1}
\end{center}
\end{table}

At this point we would like to highlight the quality of the soft-gluon approximation by comparing our approximate NLO (aNLO) results (found by adding the NLO soft-gluon corrections from Eq. (\ref{aNLO}) to the LO cross section) with the complete NLO results. At 8 TeV collision energy, the difference between the central (i.e. with $\mu=m_t$) aNLO and the NLO cross sections is entirely negligible, at the per mille level; as shown in Table 1, both results are 19.5 fb with MSHT20 NLO pdf. Varying the scale between $m_t/2$ and $2m_t$ the aNLO cross section at 8 TeV ranges from 18.8 to 20.0 fb while the NLO cross section varies from 19.5 to 19.9 fb. 

At 13 TeV energy, the difference between the central aNLO and NLO cross sections remains small, only around 3 percent. The aNLO cross section ranges from 77.4 to 82.5 fb while the NLO ranges from 77.9 to 79.3 fb. At 14 TeV energy, the aNLO cross section ranges from 94.4 to 100.6 fb and the NLO ranges from 94.5 to 96.2 fb. We note that at all three energies the NLO scale variation lies entirely within the aNLO scale variation.

Thus, at LHC energies the soft-gluon corrections account for the majority of the complete corrections, they provide very good approximations to the complete results at NLO, and they need to be considered and included at higher orders beyond NLO for more robust theoretical predictions. For reference, Table 1 shows the central cross sections at LO, aNLO, NLO, and aNNLO for 8, 13, and 14 TeV LHC energies.

\begin{figure}[htbp]
\begin{center}
\includegraphics[width=91mm]{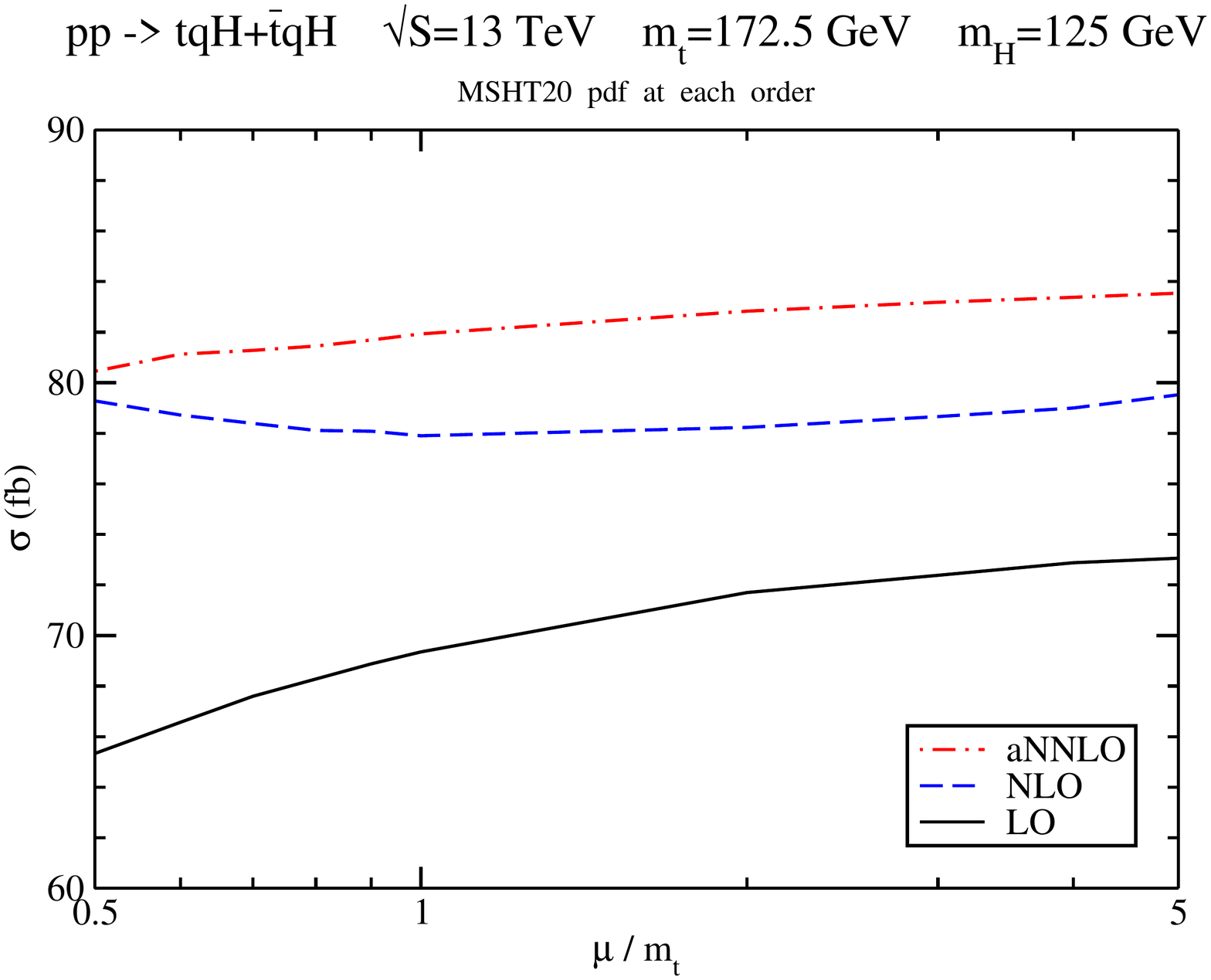}
\hspace{-7mm}
\includegraphics[width=91mm]{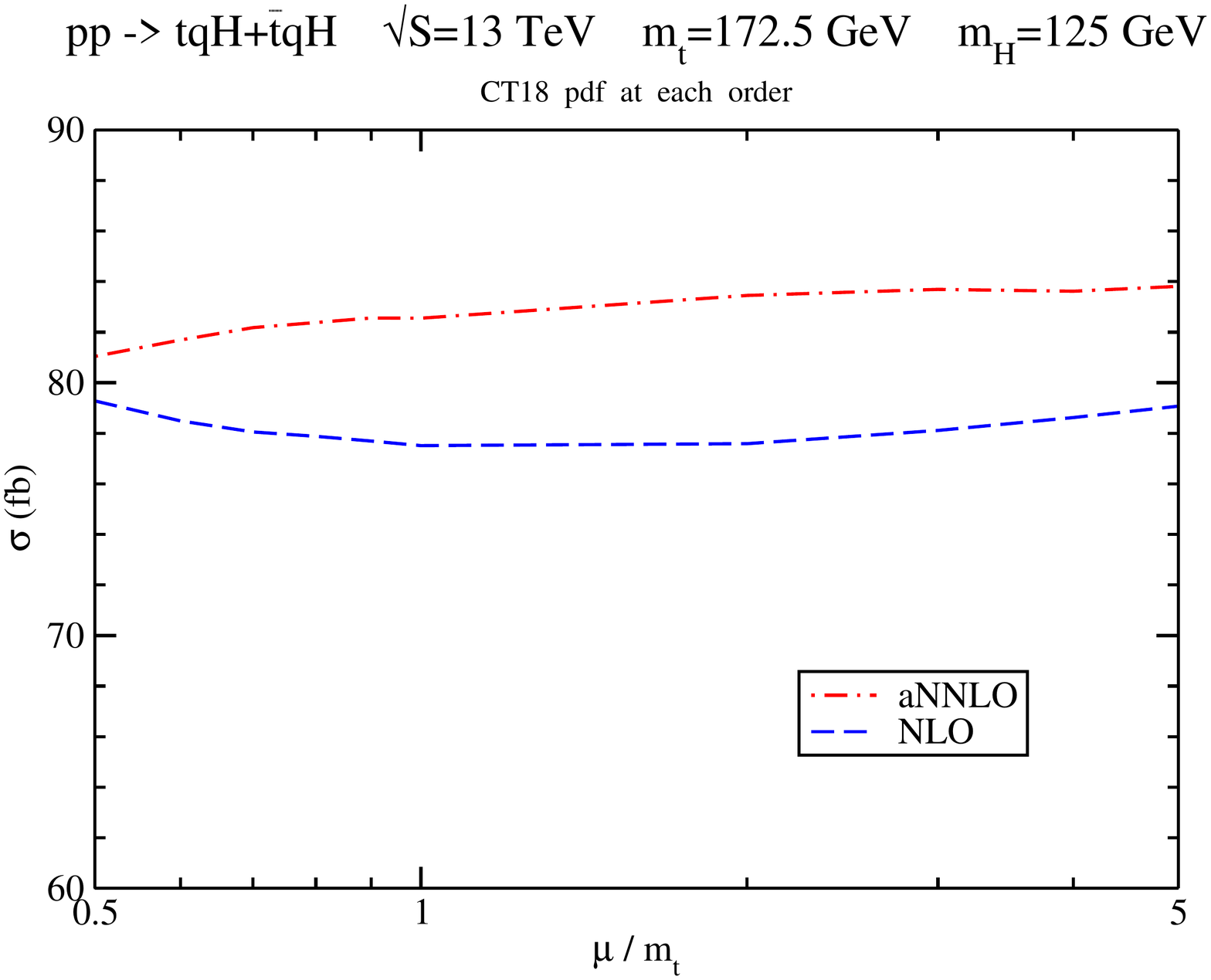}
\caption{The scale dependence of the LO, NLO, and aNNLO total cross sections for $tqH+{\bar t}qH$ production at 13 TeV LHC energy using (left) MSHT20 pdf and (right) CT18 pdf.}
\label{tqHmu13}
\end{center}
\end{figure}

In Fig. \ref{tqHmu13} we show the scale variation of the cross sections at 13 TeV LHC energy. The plot on the left shows the LO, NLO, and aNNLO results as functions of $\mu/m_t$ using MSHT20 pdf at the corresponding order. We see that while the LO cross section has a relatively large scale variation, this dependence is smaller at NLO and aNNLO. The plot on the right shows the scale variation of the NLO and aNNLO cross sections using CT18 pdf, with very similar results. We see a rather small variation at aNNLO, less than 4\%, between the maximum and minimum values over this entire range, indicating a very stable result. If we write the result with central scale $\mu=m_t$ and traditional scale variation between $m_t/2$ and $2m_t$, then the cross section at 13 TeV energy is $81.9^{+0.9}_{-1.5}$ fb at aNNLO with MSHT20 pdf. We note that the pdf uncertainty is also small, +1.2\% -0.7\% with MSHT20 pdf and +3.0\% -2.0\% with CT18 pdf.

\begin{figure}[htbp]
\begin{center}
\includegraphics[width=91mm]{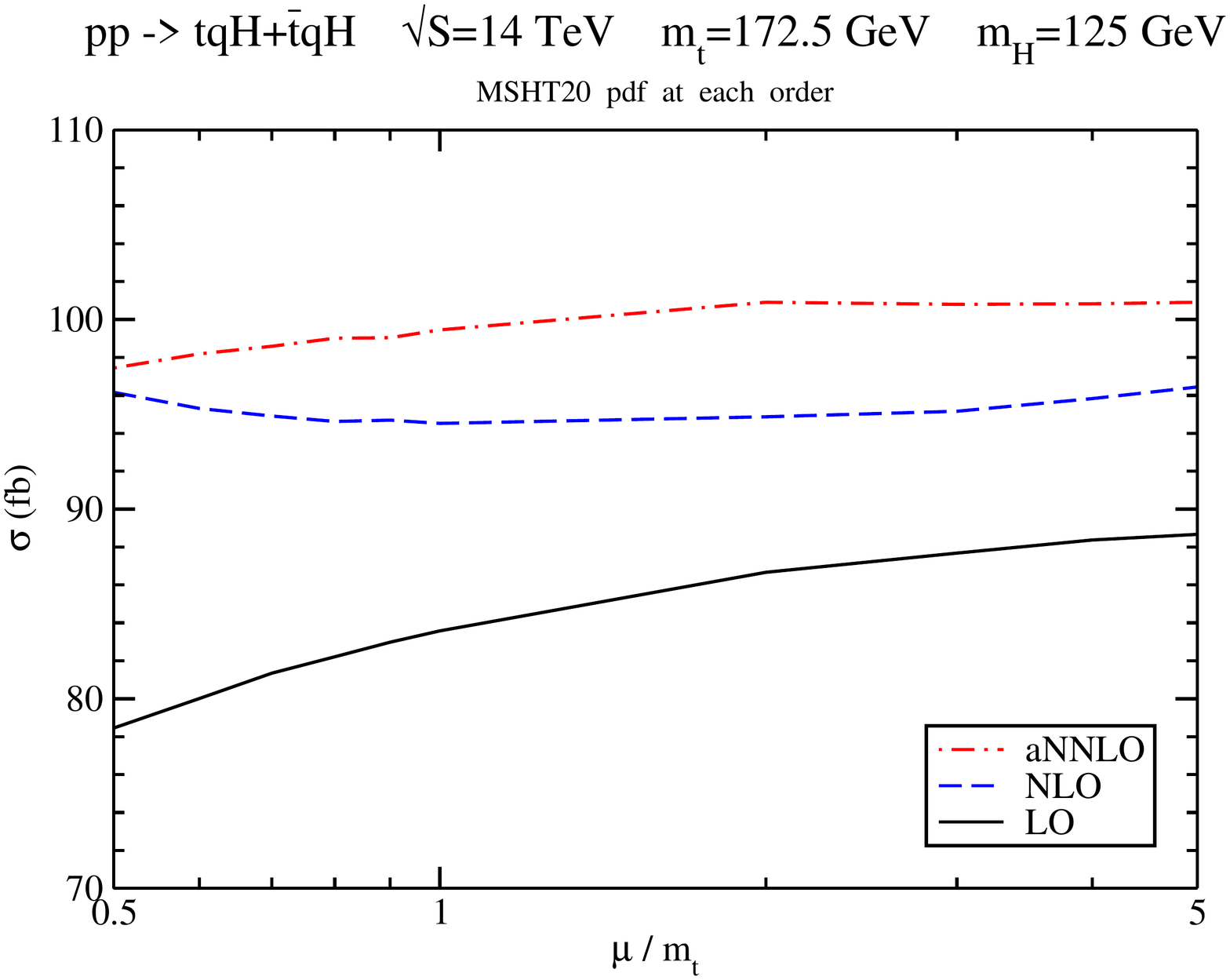}
\hspace{-7mm}
\includegraphics[width=91mm]{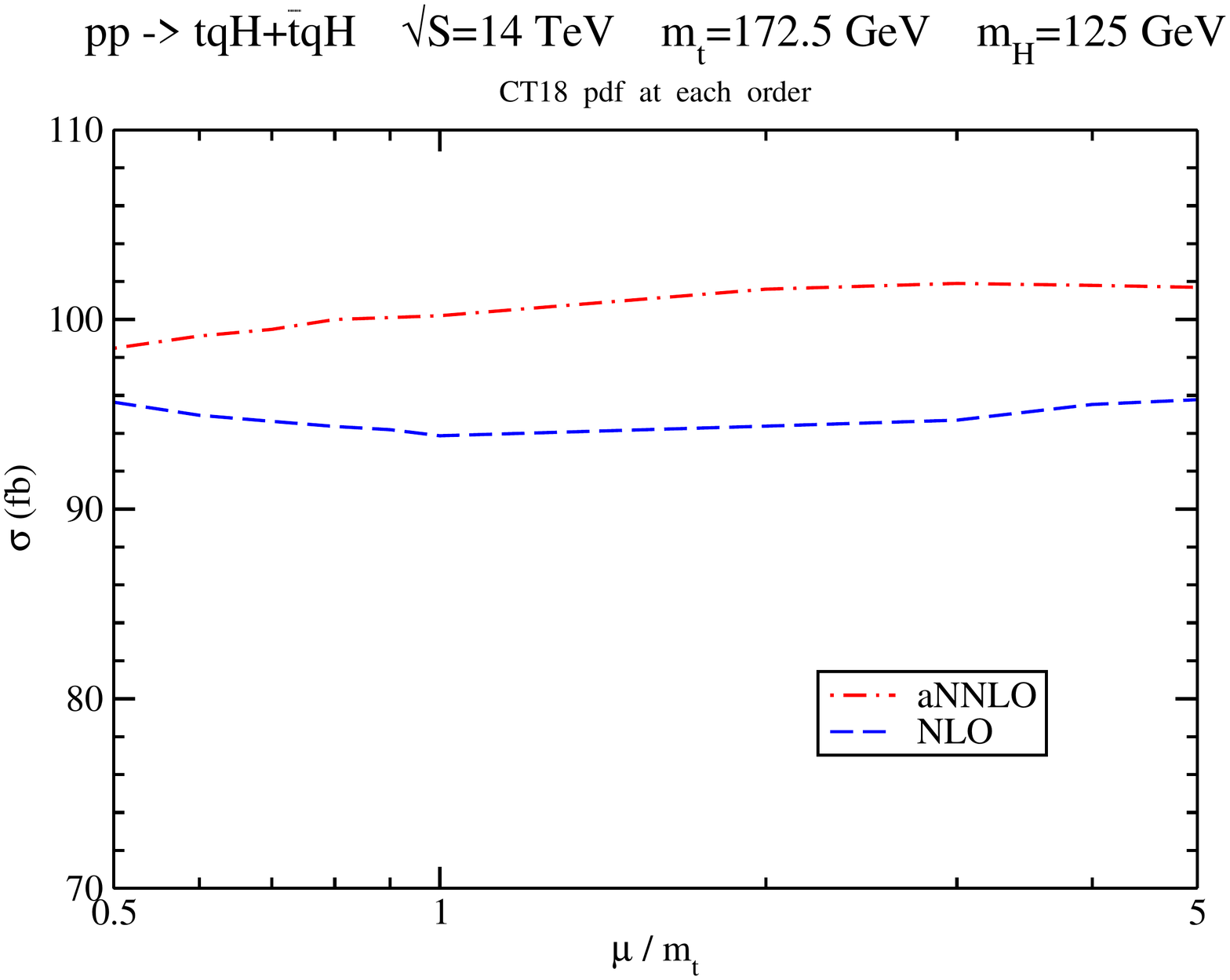}
\caption{The scale dependence of the LO, NLO, and aNNLO total cross sections for $tqH+{\bar t}qH$ production at 14 TeV LHC energy using (left) MSHT20 pdf and (right) CT18 pdf.}
\label{tqHmu14}
\end{center}
\end{figure}

In Fig. \ref{tqHmu14} we show the scale variation of the cross sections at 14 TeV LHC energy. Again, the plot on the left shows LO, NLO, and aNNLO results with MSHT20 pdf while the one on the right shows NLO and aNNLO results with CT18 pdf. The conclusions regarding the decreasing variation at higher orders relative to LO is the same as for 13 TeV. Again, we see a small variation at aNNLO of around 3.5\% between the maximum and minimum values.  If we write the result with central scale $\mu=m_t$ and traditional scale variation between $m_t/2$ and $2m_t$, then the cross section at 14 TeV energy is $99.5^{+1.4}_{-2.0}$ fb at aNNLO with MSHT20 pdf. Again, the pdf uncertainty is also small, +1.1\% -0.6\% with MSHT20 pdf and +2.8\% -1.9\% with CT18 pdf.

\subsection{Top-quark $p_T$ and rapidity distributions}

\begin{figure}[htbp]
\begin{center}
\includegraphics[width=91mm]{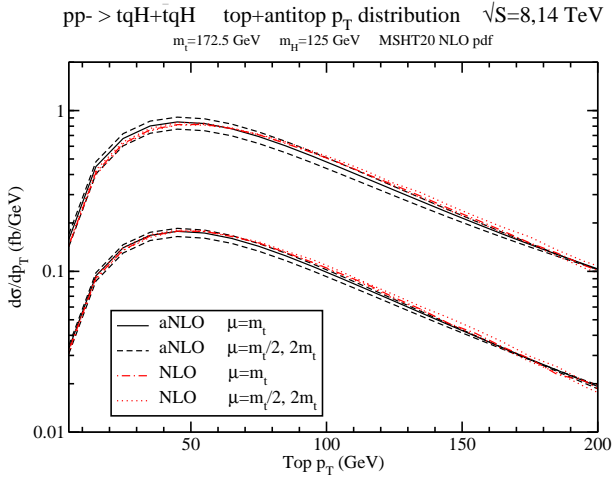}
\hspace{-7mm}
\includegraphics[width=91mm]{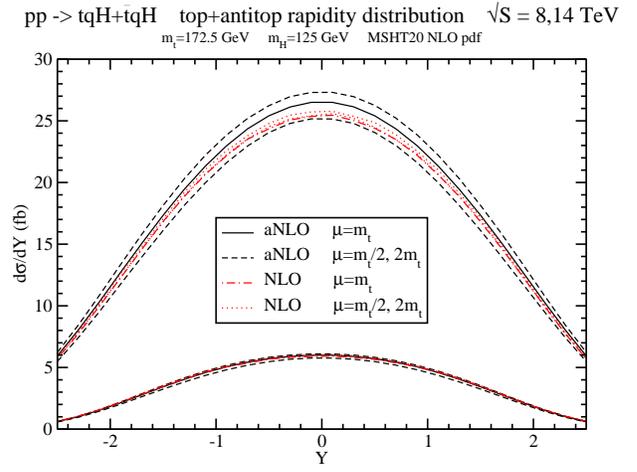}
\caption{A comparison between the aNLO and NLO results at 8 and 14 TeV energies for the (left) top-quark $p_T$ distributions and (right) rapidity distributions in $tqH+{\bar t}qH$ production.}
\label{ptytqH}
\end{center}
\end{figure}

Differential distributions provide more information than total cross sections and can be useful in the search for new physics. Transverse-momentum ($p_T$) and rapidity distributions of the top quark are typically measured in such processes. In Fig. \ref{ptytqH} we show the quality of the soft-gluon approximation for the $p_T$ distributions as well as the rapidity distributions. We compare the aNLO predictions for these quantities, including scale variation, with the corresponding NLO results at 8 and 14 TeV energies. We find that the results overlap throughout the ranges in the plots, and thus that our approximation is good not only at the level of total cross sections but also at the more detailed level of differential distributions.  

\begin{figure}[htbp]
\begin{center}
\includegraphics[width=91mm]{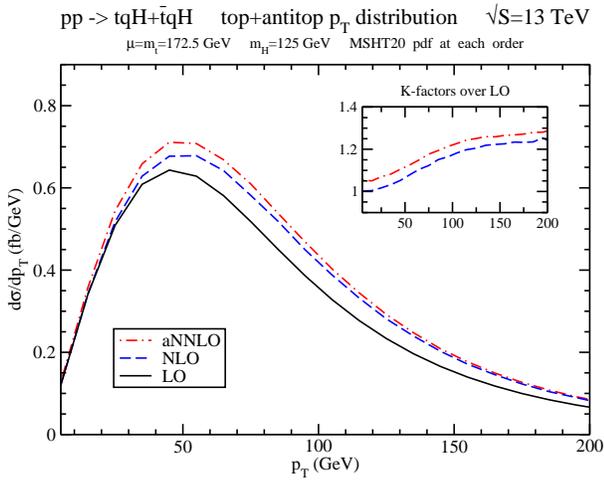}
\hspace{-7mm}
\includegraphics[width=91mm]{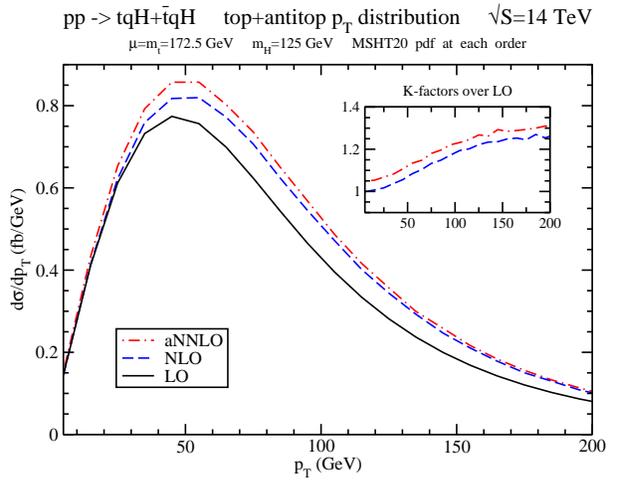}
\caption{The LO, NLO, and aNNLO top-quark $p_T$ distributions in $tqH+{\bar t}qH$ production at (left) 13 TeV and (right) 14 TeV LHC energies.}
\label{ptqH}
\end{center}
\end{figure}

In Fig. \ref{ptqH} we present the top-quark transverse-momentum distributions at 13 TeV (left) and 14 TeV (right) energies using MSHT20 pdf. We show LO, NLO, and aNNLO distributions. We observe that the aNNLO distributions peak at a $p_T$ value of around 50 GeV and quickly diminish at high $p_T$ values. The inset plots show the $K$-factors relative to LO, which indicate very significant contributions from the NLO and aNNLO corrections that increase with higher $p_T$; the combined corrections through aNNLO provide an enhancement of around 30\% at a $p_T$ of 200 GeV at both 13 and 14 TeV energies.

\begin{figure}[htbp]
\begin{center}
\includegraphics[width=91mm]{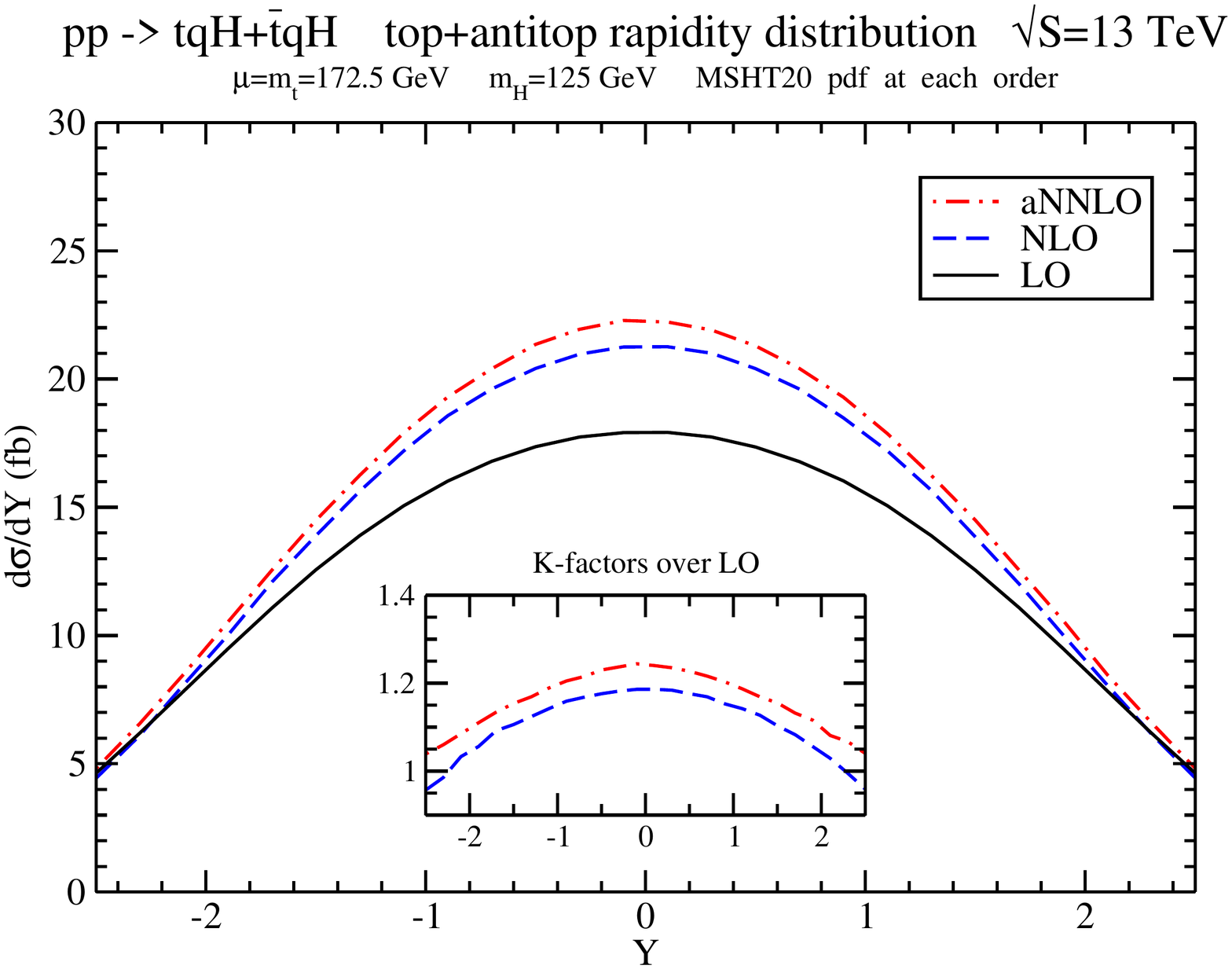}
\hspace{-7mm}
\includegraphics[width=91mm]{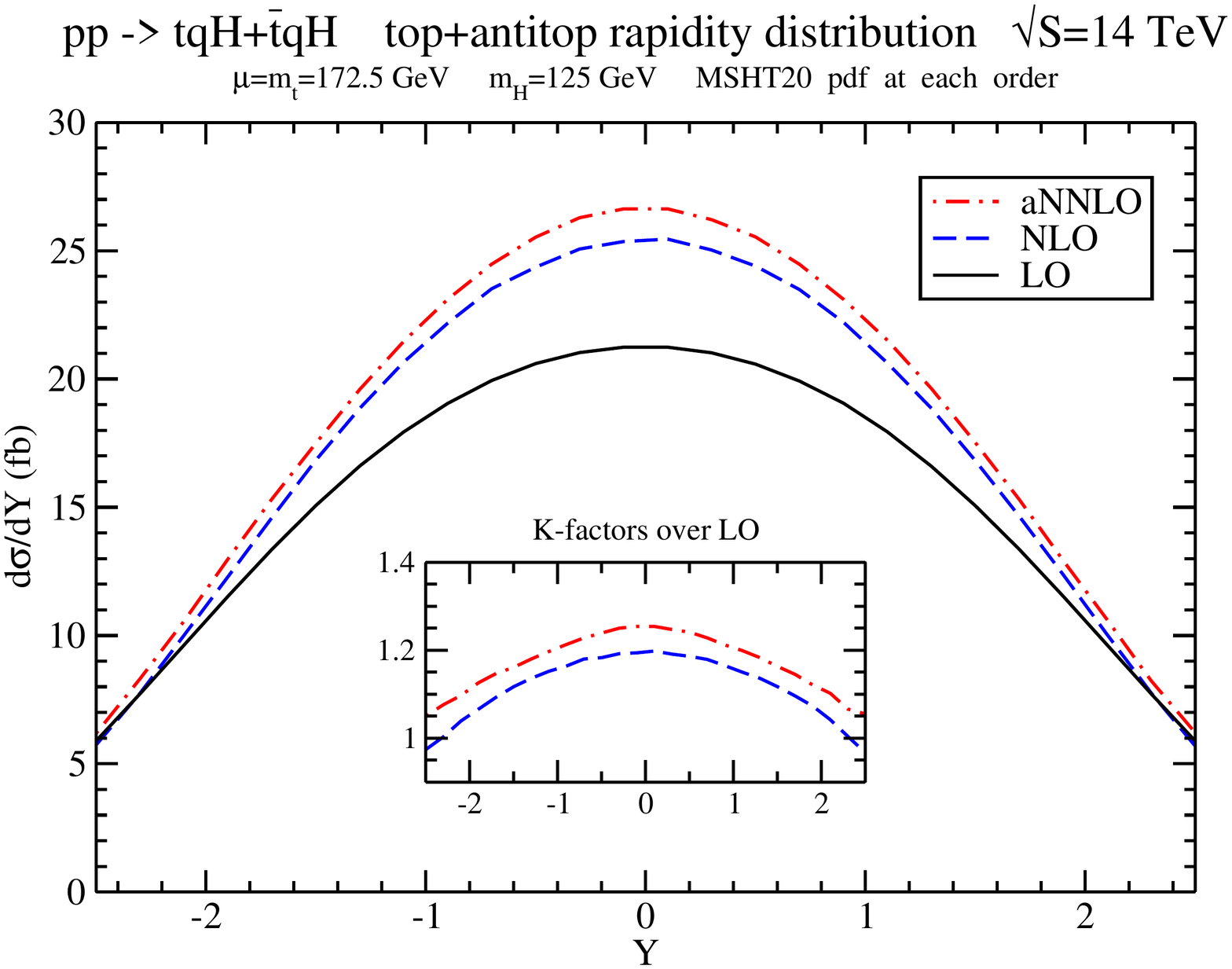}
\caption{The LO, NLO, and aNNLO top-quark rapidity distributions in $tqH+{\bar t}qH$ production at (left) 13 TeV and (right) 14 TeV LHC energies.}
\label{ytqH}
\end{center}
\end{figure}

In Fig. \ref{ytqH} we present the top-quark rapidity distributions at 13 TeV (left) and 14 TeV (right) energies using MSHT20 pdf. We show LO, NLO, and aNNLO distributions. The inset plots show the $K$-factors relative to LO, which indicate very significant contributions from the NLO and aNNLO corrections that are highest at central rapidities. The combined corrections through aNNLO provide an enhancement of around 25\% at zero rapidity.

\mysection{Conclusions}

We have presented a formalism for soft-gluon resummation for $tqH$ production in single-particle-inclusive kinematics. This is the first $2 \to 3$ process for which resummation has been performed in this kinematics. We have provided analytical results for the resummed cross section and fixed-orders expansions, and the soft anomalous dimension matrices at one and two loops. 

We have also calculated numerical results for $tqH+{\bar t}qH$ production at LHC energies. In particular, we have displayed results for the total cross sections as well as the top-quark transverse-momentum and rapidity distributions through approximate NNLO. We have shown that the soft-gluon corrections are numerically dominant and provide excellent approximations to the complete results at NLO. We have also shown that the aNNLO soft-gluon corrections are significant and their inclusion improves the theoretical predictions.

\section*{Acknowledgements}
This material is based upon work supported by the National Science Foundation under Grant No. PHY 1820795.


\begin{thebibliography}{99}

\bibitem{CMS8tev}
CMS Collaboration, V. Khachatryan {\it et al.}, JHEP {\bf 06}, 177 (2016) [arXiv:1509.08159].

\bibitem{CMS13tev}
CMS Collaboration, A. M. Sirunyan {\it et al.}, Phys. Rev. D {\bf 99}, 092005 (2019) [arXiv:1811.09696].

\bibitem{CER}
J. Campbell, R.K. Ellis, and R. Rontsch, Phys. Rev. D {\bf 87}, 114006 (2013) [arXiv:1302.3856].

\bibitem{DMMZ}
F. Demartin, F. Maltoni, K. Mawatari, and M. Zaro, Eur. Phys. J. C {\bf 75}, 267 (2015) [arXiv:1504.00611].

\bibitem{NKGS}
N. Kidonakis and G. Sterman,  Phys. Lett. B {\bf 387}, 867 (1996); Nucl. Phys. B {\bf 505}, 321 (1997) [hep-ph/9705234].

\bibitem{CLS}
H. Contopanagos, E. Laenen, and G. Sterman, Nucl. Phys. B {\bf 484}, 303 (1997) [hep-ph/9604313]. 

\bibitem{KOS}
N. Kidonakis, G. Oderda, and G. Sterman, Nucl. Phys. B {\bf 525}, 299 (1998) [hep-ph/9801268]; Nucl. Phys. B {\bf 531}, 365 (1998) [hep-ph/9803241].

\bibitem{LOS}
E. Laenen, G. Oderda, and G. Sterman, Phys. Lett. B {\bf 438}, 173 (1998) [hep-ph/9806467].

\bibitem{HRSV}
P. Hinderer, F. Ringer, G. Sterman, and W. Vogelsang, Phys. Rev. D {\bf 99}, 054019 (2019) [arXiv:1812.00915].

\bibitem{FK2020}
M. Forslund and N. Kidonakis, Phys. Rev. D {\bf 102}, 034006 (2020) [arXiv:2003.09021].

\bibitem{NKtoprev}
N. Kidonakis, Int. J. Mod. Phys. A {\bf 33}, 1830021 (2018) [arXiv:1806.03336].

\bibitem{NKn3lo} 
N. Kidonakis, Phys. Rev. D {\bf 90}, 014006 (2014) [arXiv:1405.7046];
D {\bf 91}, 031501 (2015) [arXiv:1411.2633]; 
D {\bf 91}, 071502 (2015) [arXiv:1501.01581];
D {\bf 101}, 074006 (2020) [arXiv:1912.10362].

\bibitem{NKst}
N. Kidonakis, Phys. Rev. D {\bf 81}, 054028 (2010) [arXiv:1001.5034]; 
D {\bf 82}, 054018 (2010) [arXiv:1005.4451]; D {\bf 83}, 091503 (2011) [arXiv:1103.2792]; D {\bf 96}, 034014 (2017) [arXiv:1612.06426].

\bibitem{NKNY}
N. Kidonakis and N. Yamanaka, JHEP {\bf 05}, 278 (2021) [arXiv:2102.11300].

\bibitem{NKtH}
N. Kidonakis, Phys. Rev. D {\bf 94}, 014010 (2016) [arXiv:1605.00622]. 

\bibitem{NKtZ}
N. Kidonakis, Phys. Rev. D {\bf 97}, 034028 (2018) [arXiv:1712.01144]. 

\bibitem{MFNK}
M. Forslund and N. Kidonakis, Phys. Rev. D {\bf 98}, 074017 (2018) [arXiv:1808.09014].

\bibitem{MGNK}
M. Guzzi and N. Kidonakis, Eur. Phys. J. C {\bf 80}, 467 (2020) [arXiv:1904.10071]. 

\bibitem{NK2020rev}
N. Kidonakis, Universe {\bf 6}, 165 (2020) [arXiv:2008.09914].

\bibitem{LLL}
H.T. Li, C.S. Li, and S.A. Li, Phys. Rev. D {\bf 90}, 094009 (2014) [arXiv:1409.1460].

\bibitem{KMST}
A. Kulesza, L. Motyka, T. Stebel, and V. Theeuwes, 
JHEP {\bf 03}, 065 (2016) [arXiv:1509.02780]. 

\bibitem{BFPSY}
A. Broggio, A. Ferroglia, B.D. Pecjak, A. Signer, and L.L. Yang, 
JHEP {\bf 03}, 124 (2016) [arXiv:1510.01914].

\bibitem{BFOP}
A. Broggio, A. Ferroglia, G. Ossola, and B.D. Pecjak, 
JHEP {\bf 09}, 089 (2016) [arXiv:1607.05303].

\bibitem{BFPY}
A. Broggio, A. Ferroglia, B.D. Pecjak, and L.L. Yang, 
JHEP {\bf 02}, 126 (2017) [arXiv:1611.00049].

\bibitem{BFOPS}
A. Broggio, A. Ferroglia, G. Ossola, B.D. Pecjak, and R.S. Sameshima, 
JHEP {\bf 04}, 105 (2017) [arXiv:1702.00800].

\bibitem{KMST2}
A. Kulesza, L. Motyka, T. Stebel, and V. Theeuwes, 
Phys. Rev. D {\bf 97}, 114007 (2018) [arXiv:1704.03363].

\bibitem{KMSST}
A. Kulesza, L. Motyka, D. Schwartlander, T. Stebel, and V. Theeuwes, 
Eur. Phys. J. C {\bf 79}, 249 (2019) [arXiv:1812.08622].

\bibitem{BFFPPT}
A. Broggio, A. Ferroglia, R. Frederix, D. Pagani, B.D. Pecjak, and I. Tsinikos, 
JHEP {\bf 08}, 039 (2019) [arXiv:1907.04343]. 

\bibitem{GS}
G. Sterman, Nucl. Phys. B {\bf 281}, 310 (1987). 

\bibitem{ADS}
S.M. Aybat, L.J. Dixon, and G. Sterman, Phys. Rev. Lett. {\bf 97}, 072001 (2006) [arXiv:hep-ph/0606254].

\bibitem{NK2loop}
N. Kidonakis, Phys. Rev. Lett. {\bf 102}, 232003 (2009) [arXiv:0903.2561]; Phys. Rev. D {\bf 82}, 114030 (2010) [arXiv:1009.4935].

\bibitem{NK3loop}
N. Kidonakis, Int. J. Mod. Phys. A {\bf 31}, 1650076 (2016) [arXiv:1601.01666];  Phys. Rev. D {\bf 99}, 074024 (2019) [arXiv:1901.09928].

\bibitem{CT}
S. Catani and L. Trentadue, Nucl. Phys. B {\bf 327}, 323 (1989).

\bibitem{MSHT20}
S. Bailey, T. Cridge, L.A. Harland-Lang, A.D. Martin, and R.S. Thorne, Eur. Phys. J. C {\bf 81}, 341 (2021) [arXiv:2012.04684].

\bibitem{CT18}
T.-J. Hou, J. Gao, T.J. Hobbs, K. Xie, S. Dulat, M. Guzzi, J. Huston, P. Nadolsky, J. Pumplin, C. Schmidt, I. Sitiwaldi, D. Stump, and C.-P. Yuan, Phys. Rev. D {\bf 103}, 014013 (2021) [arXiv: 1912.10053].

\bibitem{lhapdf}
A. Buckley {\it et al.}, Eur. Phys. J. C {\bf 75}, 132 (2015) [arXiv:1412.7420].

\bibitem{MG5}
J. Alwall {\it et al.}, JHEP {\bf 07}, 079 (2014) [arXiv:1405.0301].

\end{thebibliography}
\end{document}